\documentclass{article}

\usepackage{float}
\usepackage{arxiv}
\usepackage[utf8]{inputenc} 
\usepackage[T1]{fontenc}    
\usepackage{url}            
\usepackage{booktabs}       
\usepackage{amsfonts}       
\usepackage{nicefrac}       
\usepackage{microtype}      
\usepackage{lipsum}
\usepackage{amsmath}
\usepackage{graphicx}
\usepackage{color}
\usepackage{comment}
\usepackage{setspace}
\title{Accelerating Adjoint Variable Method Based Photonic Optimization with Schur Complement Domain Decomposition}

\author{
  Nathan Zhao \\
  Department of Applied Physics\\
  Stanford University\\
  Stanford, CA 94305 \\
  \texttt{nzz2102@stanford.edu} \\
   \And
  Salim Boutami \\
  Universite Grenoble Alpes\\
  CEA, LETI  \\
  Grenoble, 38054, France\\
   \AND
  Shanhui Fan\\
  Department of Electrical Engineering\\
  Stanford University \\
  Stanford, CA 94305 \\
   \texttt{shanhui@stanford.edu} \\
}

\begin{document}
\maketitle

\begin{abstract}
Adjoint variable method in combination with gradient descent optimization has been widely used for the inverse design of nanophotonic devices. In many of such optimizations, the design region is only a small fraction of the total computational domain. Here we show that the adjoint variable method can be combined with the Schur complement domain decomposition method. With this combination, in each optimization step, the simulation only involves the degrees of freedom that are inside the design region. Our approach should significantly improve the computational efficiency of adjoint variable method based optimization of photonic structures. 
\end{abstract}

\section{Introduction}
The adjoint variable method has been well established for inverse design for passive and linear devices in nanophotonics \citep{Piggott2015, Liu2011, Lalau-Keraly2013, Veronis2004, Georgieva2002, Lu2012}, and has been recently generalized to active and nonlinear systems \citep{Wang2018, Hughes2018}. In particular, the technique has the computational advantage of requiring exactly two full-field simulations in order to determine the gradient of an objective with respect to an arbitrarily large number of design parameters in the domain. Combined with a gradient descent search, the method significantly decreases the number of structures that need to be probed in a given design problem. 

In many optical design problems, for a given structure, one typically optimizes only part of structure (referred to as the ``design region"), while the rest of the structure (referred to as the ``background") is kept unchanged. (Fig. 1). For many problems, this design region can often be a very small fraction of the entire computational domain \citep{Estakhri2019}. For the standard implementation of adjoint variable method based gradient-descent optimization, in each step of the optimization, one simulates the entire structure including the background, even though the background is unchanged between the optimization steps. In order to reduce the computational cost, it would certainly be advantageous if in each optimization step one instead only performs the simulation on the design region.


In this context, the technique of a Schur complement domain decomposition is of interest. The techniques of domain decomposition have been widely used to accelerate large-scale simulations by parallelization \citep{Dolean2015} as well as to design nested dissection algorithms for direct solvers for sparse linear systems \citep{George1973, Conroy1990}. In photonic device design, Ref. \citep{Verweij2014} has shown that the Schur-complement domain-decomposition technique can be fruitfully applied in a discrete optimization scheme. However, in photonics, there has not been any work in applying the Schur complement domain-decomposition technique in continuous optimization schemes such as those based on the adjoint variable methods. 


In this paper, we detail how the Schur complement domain decomposition approach can be combined with adjoint variable optimization in photonics device design. Since the adjoint variable method is generally applicable to any linear systems, and the Schur complement domain decomposition method describes the photonic device of interest as linear systems, it should not come as a surprise that one can combine these two methods. However, there are subtleties in the combination. In particular, in many photonic device design problem, the excitation source, as well as the field components that define the figure of merits, are not in the design region. We show that the formalism of the Schur complement naturally deals with such situations. The proposed method here, by being able to selectively simulate only the subdomain of interest, points to an avenue for increasing the computational efficiency of adjoint-based optimizations in photonics. 

The rest of the paper is organized as follows. In Section 2, we outline the mathematical formalism. In Section 3, we demonstrate the computational gains of our proposed method through the design example of a mode converter. We conclude in Section 4. 

\section{Mathematical description of the algorithm}

\begin{figure}[H]
    \centering
    \includegraphics[width = 4.2 in]{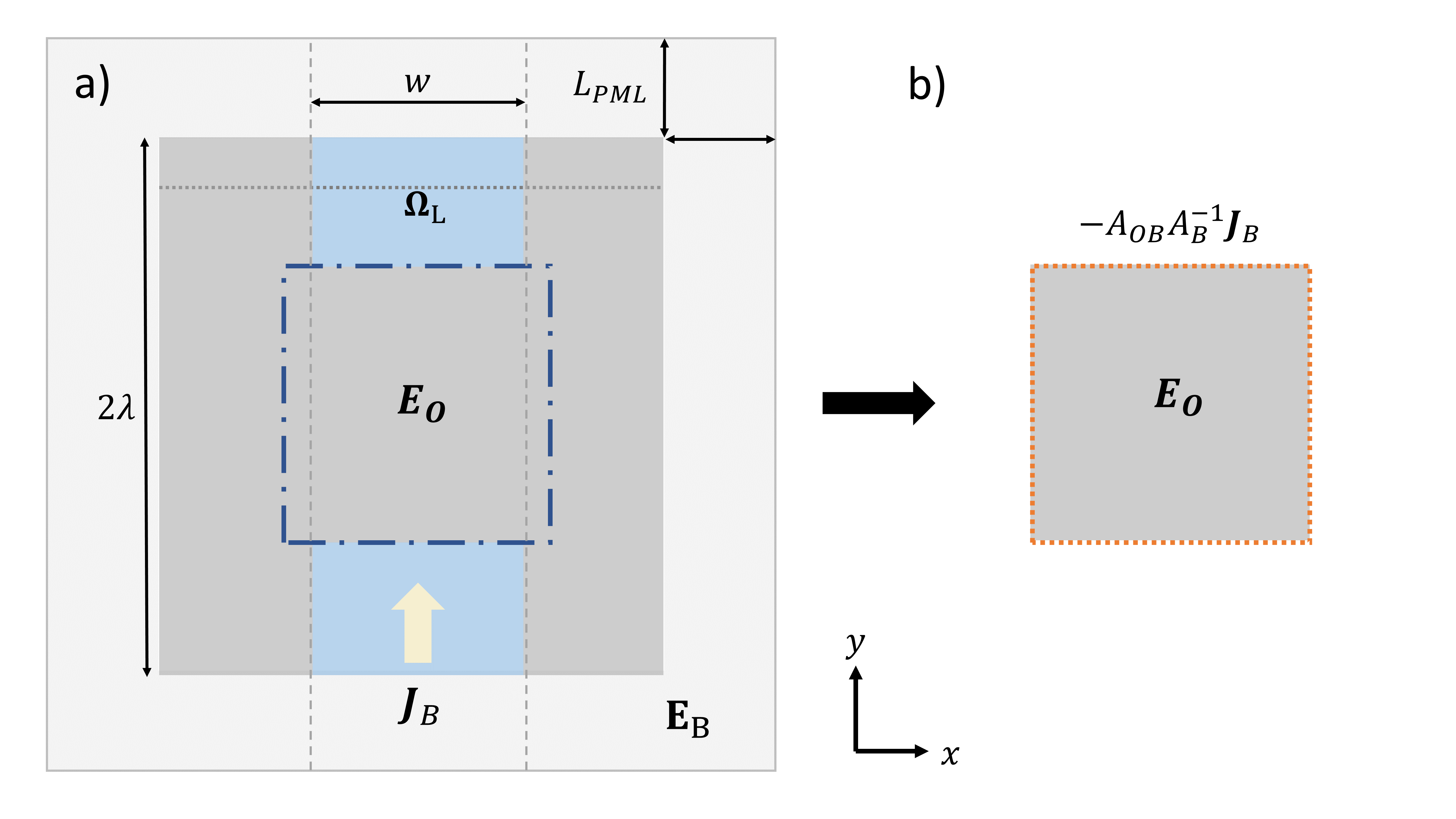}
    \caption{a)	The original system used for the design of a waveguide mode converter. The gray region has a relative permittivity $\epsilon_B = 2.25$, the light blue region has a permittivity of $\epsilon = 6.25$. The light gray region consists of perfectly matched layer. $J_B$ indicates the position of the source. The dashed line denoted by $\Omega_L$ indicates the line segment where the figure of merit is calculated. The square bounded by the dotted dashed line is the design region. b) The reduced system consisting only of the design region. The source in the original system that is outside the design region is mapped to a source on the boundary of the design region. }
    \label{fig:intro}
\end{figure}

In this section we discuss the approach for combining the method of adjoint variable with the method of Schur complement domain decomposition. First, we briefly outline the Schur complement domain decomposition as applied to a numerically discretized solution of Maxwell's equations in the frequency domain. The Maxwell's equation written for the $\mathbf{E}$-field is:
\begin{equation}
\bigg(\frac{1}{\mu_0}\nabla \times \nabla \times  - \omega^2 \epsilon_0 \epsilon_r(\mathbf{r}) \bigg)\mathbf{E(r)} = -i\omega \mathbf{J(r)}
\label{eq:maxwell}
\end{equation}
where $\mathbf{E(r)}$ denotes the vector electric field, $\mathbf{J(r)}$ is the source, $\epsilon_r(\mathbf{r})$ is the dielectric permittivity and the magnetic permeability is assumed to be $\mu_0$ in the entire computational domain. In Eq. \eqref{eq:maxwell}, $\mathbf{E(r)}$ and $\mathbf{J(r)}$ are vector complex functions.

By discretizing Eq. \eqref{eq:maxwell} via finite differences on a Yee grid, we can convert this into a system of linear equations:
\begin{equation}
    A\mathbf{E} = -i\omega \mathbf{J} 
    \label{eq:forward}
\end{equation}
where $A$ represents discretized approximation of the left hand side of Eq. \eqref{eq:maxwell}. The $\mathbf{E}$ in Eq. \eqref{eq:forward} is now a vector contains the field components, which are the unknowns, for each spatial grid point. $\mathbf{J}$ is a vector containing information about the sources on this discretized domain.

We now consider the example domain shown in Fig. \ref{fig:intro}a), which we will also use in the numerical example in the next section. The domain consists of the design region, as denoted by subscript $O$ since this is the region where the optimization takes place, and the background region, denoted by the subscript $B$, that contains the rest of the computational domain. The background in this case contains a perfectly matched layer (light gray), a lower and upper waveguide section (light blue). The input source is in the background region.  By reordering the spatial degrees of freedom in $\mathbf{E}$ and appropriately permuting the rows of $A$ in Eq. \eqref{eq:forward}, we can write Eq. \eqref{eq:forward} in a $2 \times 2$ block form:



\singlespacing
\begin{equation}
    \begin{bmatrix}
   A_O  & A_{OB} \\
   A_{BO} &  A_B
   \end{bmatrix}\begin{bmatrix}
   \mathbf{E}_O \\
   \mathbf{E}_B
   \end{bmatrix} = -i\omega \begin{bmatrix}
   \mathbf{J}_O \\
   \mathbf{J}_B
   \end{bmatrix}
   \label{eq:partitioned_A}
\end{equation}

The submatrices $A_O$ and $A_B$ are now the discretized representations of Eq. \eqref{eq:maxwell} restricted to the design region and the background, respectively. The off-diagonal blocks $A_{OB}$ and $A_{BO}$ represent the spatial couplings between grid points of these two regions. The coupling occurs along their borders. 

Using the second row of Eq. \eqref{eq:partitioned_A}
\begin{equation}
     A_{BO}\mathbf{E}_O + A_B\mathbf{E}_B = -i\omega \mathbf{J}_B
     \label{eq:second_row}
\end{equation}
we can solve for $\mathbf{E}_B$ and substitute it into the linear equation in the first row to get the equation for $\mathbf{E}_O$ only:
\begin{equation}
    S\mathbf{\mathbf{E}_O} = -i\omega \mathbf{J}_S
    \label{eq:reduced}
\end{equation}
where:
\begin{equation}
S \equiv A_O - A_{OB}A_B^{-1}A_{BO}
\label{eq:schur}
\end{equation}
\begin{equation}
    \mathbf{J}_S \equiv \mathbf{J}_O - A_{OB}A_B^{-1}\mathbf{J}_B. 
    \label{eq:schur_source}
\end{equation}
Below, we will refer to Eq. \eqref{eq:partitioned_A} as the ``original'' system and Eq. \eqref{eq:schur} as its corresponding ``reduced'' system.  From Eq. \eqref{eq:schur_source}, we see that the sources outside the design region in the original system are now represented in the reduced system by sources on the surface of the reduced computational domain, which contains only the design region. The solution to the reduced system of Eq. \eqref{eq:reduced} now contains only the degrees of freedom in $\mathbf{E}_O$. The most expensive step in forming Eq. (4) is to compute  $A_{OB} A_{B}^{-1} A_{BO}$, and $A_{OB} A_{B}^{-1} \mathbf{J}_B$. This can be done by a direct factorization of $A_B$, or by solving for $A_B^{-1} A_{BO}$ and $A_B^{-1} \mathbf{J}_B$ using a linear system solver. Importantly, this step, while expensive, only needs to be done once for the entire optimization process.  Moreover, if the background region has significant repeated structure or is highly uniform, this operation can be accelerated by breaking it into smaller subdomains and solving $A_{OB}A_B^{-1}A_{BO}$ in parallel \citep{Dolean2015}.


Finally, the solution field $\mathbf{E}_B$ in the background region can be recovered by simple substitution of the solution $\mathbf{E}_O$ into Eq. \eqref{eq:second_row}, yielding:
\begin{equation}
\mathbf{E}_B = A_B^{-1} (\mathbf{J}_O - A_{BO} \mathbf{E}_O).
\label{eq:reconstruct}
\end{equation}

In many optimization problems, the figure of merit is determined based on fields that lie outside the design region. Based on Eq. \eqref{eq:reconstruct}, such figure of merit can be straightforwardly evaluated using fields and sources inside the design region only. This observation is important when we discuss the combination with the adjoint variable method below.





We now briefly review the adjoint variable method. In adjoint method, the goal is to maximize a figure of merit  as described by a scalar objective function $L$ with respect to all the design variables $\phi$ \citep{Bendsoe2004}. Thus, it is of critical importance to compute the sensitivity of $L$  with respect to each element of $\phi$, (i.e. to compute $\partial L/\partial \phi$). Here for notation simplicity we have suppressed all indices for the elements of $\phi$. $\partial L/\partial \phi$ for example is a vector with a dimension equal to the number of design variables. To proceed we note that:
\begin{equation}
    \frac{\partial L}{\partial \phi} = \frac{\partial L}{\partial \mathbf{E}}\frac{\partial \mathbf{E}}{\partial \phi}
    \label{eq:sensitivity}
\end{equation}
$\frac{\partial L}{\mathbf{\partial E}}$ can be determined analytically since the objective function $L$ is typically an analytic function of $\mathbf{E}$. To determine the second term $\partial \mathbf{E}/ \partial \phi$, we take derivatives of Eq. \eqref{eq:forward} with respect to $\phi$ to obtain:
\begin{equation}
    \frac{\partial \mathbf{E}}{\partial \phi} = -A^{-1} \frac{\partial A}{\partial \phi} \mathbf{E}
    \label{eq:deriv_Ae}
\end{equation}
With a simple substitution of Eq. \eqref{eq:deriv_Ae} into Eq. \eqref{eq:sensitivity}, we get:
\begin{equation}
    \frac{\partial L}{\partial \phi} = -\bigg(\frac{\partial L}{\partial \mathbf{E}} A^{-1} \bigg) \frac{\partial A}{\partial \phi} \mathbf{E} = - \hat{\mathbf{E}}^T\frac{\partial A}{\partial \phi}\mathbf{E}
\end{equation}
The term $\hat{\mathbf{E}}$ is the adjoint field and can be determined by solving:
\begin{equation}
    A^T \hat{\mathbf{E}} = \bigg(\frac{\partial L}{\partial \mathbf{E}}\bigg)^T
    \label{eq:adjoint}
\end{equation}
where the term on the right hand side is the adjoint source. The combined solutions of $\mathbf{E}$ and $\hat{\mathbf{E}}$ are enough to compute the gradient of the objective function $L$ with respect to all design variables:
\begin{equation}
    \frac{\partial L}{\partial \phi} = -\hat{\mathbf{E}}^T \frac{\partial A}{\partial \phi} \mathbf{E}
\end{equation}

In the adjoint-variable-method based optimization, at each step of the optimization, we update or modify the design variables $\phi$ according the sensitivity information $\partial L/\partial \phi$:
\begin{equation}
    \phi_{new} = \phi_{old} +\alpha\bigg(\frac{\partial L}{\partial \phi}\bigg)
    \label{eq:grad_update}
\end{equation}
This ensures that we modify the structure in a locally optimal way for the improvement of the objective function. $\alpha$ is the learning rate. Heuristic ways of choosing or adapting $\alpha$ can be found in Ref. \citep{Ruder2016}.

In combining the adjoint variable method with the Schur complement domain decomposition method, the steps in Eqs. \eqref{eq:sensitivity} - \eqref{eq:grad_update} remain the same, but we now apply it to Eq. \eqref{eq:reduced} instead of Eq. \eqref{eq:forward}. In this case, the adjoint sensitivity can be determined by computing for both $\mathbf{E}_O$ as well as the adjoint field $\hat{\mathbf{E}}_O^T$:
\begin{equation}
     \frac{\partial L}{\partial \phi} = -\hat{\mathbf{E}}_O^T \frac{\partial S}{\partial \phi} \mathbf{E}_O
    \label{eq:schur_sensitivity}
\end{equation}
In that way, simulations are done only on the degrees of freedom inside the design region (i.e. directly solving the linear system in Eq. \eqref{eq:reduced}. During the optimization process, from Eq. \eqref{eq:reduced}, the only term in $S$ that changes between each optimization step is $A_O$. Consequently, the term $A_{OB}A_B^{-1}A_{BO}$ can be precomputed once and re-used throughout the optimization process. 

In many optical design problems, the objective function is defined using the field inside the background region. As an example, 
\begin{equation}
    L \propto \mathbf{E}_B^{\dagger} (\hat M \mathbf{E}_B)
\end{equation}
where $\hat M$ is a linear operator. In order to evaluate such cost function using the Schur complement domain decomposition approach, where only the field inside the design region is specifically simulated, we note that the adjoint source corresponding to this is:
\begin{equation}
    \mathbf{J}_B = \frac{\partial L}{\partial \mathbf{E}_B} = \hat M \mathbf{E}_B
    \label{eq:adjoint_source}
\end{equation}
where we use the subscript $B$ to signify that the relevant field, and the source distribution, are located in the background region and are outside the design region. The corresponding source, for the reduced system, can then be determined using Eq. \eqref{eq:schur_source}.  This adjoint source in the original system thus is mapped to a source in the reduced system. The computation required for this mapping only needs to be carried out once in the beginning of the optimization process. Thus, we have shown the adjoint variable method can be readily combined with the Schur complement domain decomposition method, including for the cases where the figure of merit involve fields that are outside the design region.

\section{Application to Accelerating the Design of a Mode Converter}
In this section, we demonstrate the performance gains of applying our formalism in the exemplary problem of designing a mode converter on the domain shown in Fig. \ref{fig:intro}a). For simplicity, we restrict our study to two-dimensional structures (i.e. structures that are infinite and translationally invariant in the out-of-plane $z$-direction) and study the TE polarization ($E_z$, $H_x$, $H_y$).  In this example, the design variable $\phi$ is the spatial dielectric permittivity distribution $\epsilon_r(x,y)$.

The objective of the device design, as shown in Fig. \ref{fig:intro}a), will be to transmit the TE$_0$ mode of an input waveguide to the TE$_1$-mode of the output waveguide with minimal reflections. For this purpose we will optimize for the dielectric function permittivity distribution in the design region between the two waveguides, as shown in Fig. \ref{fig:intro}a). Typical mode converters in dielectric waveguides require relatively long structures (on the order of several wavelengths) \citep{Chen2015}. Designing a compact version of this device would be very useful. 

To simulate this device, the computational domain, as shown in Fig. \ref{fig:intro}a), contains a square region with an edge length of $2\lambda$ with $\lambda = 1.55$ $\mu$m. The square region contains the input and output waveguides, shown as sky blue. The waveguides have a guiding layer with relative permittivity $\epsilon$ = 6.25 and width of $w=1$ $\mu$m, surrounded by a background material with a permittivity $\epsilon_B = 2.25$. Between the input and output waveguides is the design region, which is also a square. In the optimization process, the initial structure in the design region is assumed to be uniform with a relative dielectric permittivity of 2.25. The square region is surrounded on all sides by a perfectly matched layer (PML) \citep{Berenger1994} with a thickness $L_{PML}$ of 0.75 $\mu$m. The entire computational domain is discretized on a square lattice grid with a grid spacing of 50 nm. 


The figure of merit $L$ for this mode converter is proportional to the overlap integral involving the electric field along a line perpendicular to the output waveguide, and the spatial profile of the TE$_1$ magnetic field mode of the output waveguide \citep{Lalau-Keraly2013}:
\begin{equation}
   L = \bigg|\int_{\Omega_{L}}{E_{z}(x)H_{x,TE_1}^*(x)+ E_{z,TE_1}^*(x)H_{x}(x) dx \bigg|^2}
    \label{eq:FoMwaveguide}
\end{equation}
where $\Omega_L$ denotes the line along which we integrate. The derivative of this figure of merit with respective to the dielectric permittivity is $\epsilon_r$ is:
\begin{equation}
    \frac{\partial L}{\partial \epsilon_r} = 2 k_0^2 \text{Re}\bigg(E_z(x)\hat{E}_z(x) L^*\bigg)
\end{equation}
To search for the structure which optimizes the figure of merit, we perform gradient descent optimizations for 450 iterations using both Eq. \eqref{eq:partitioned_A} and Eq. \eqref{eq:reduced} and compare the total time taken to optimize the device. To solve these equations, we use an LU factorization \citep{Davis2004}. To solve for the adjoint field, we then use back-substitution of the LU factorization, which is considerably faster. The number of iterations is chosen to be sufficiently large so that the figures of merit converge at the end of the iteration processes. In the gradient descent optimization, we gradually reduce the learning rate $\alpha$ in Eq. \eqref{eq:grad_update} as the optimization progresses.  We denote the total time to optimize the device using the reduced system of Eq. \eqref{eq:reduced} as $T_S$. $T_S$ does not includes the overhead cost $T_0$ of evaluating $A_{OB} A_{B}^{-1} A_{BO}$ and $A_{OB} A_{B}^{-1} \mathbf{J}_B$ in Eq. \eqref{eq:reduced}. We denote the total time to optimize the device using the original system of Eq. \eqref{eq:partitioned_A} as $T_A$.

We compare the performance of solving for either original or the reduced systems, in two different optimization approaches. In the first approach, which we will refer to as the "permittivity" approach, we vary the permittivity in each pixel continuously according the gradient of the cost function with respect to the permittivity in each pixel, while restricting the permittivity of each pixel to be within the range of [2.25, 6.25]. The resulting optimal structure then can have any real value between $[2.25, 6.25]$ in the design region. The second approach, which will be referred to as the "level-set" approach, involves using a level set technique so that the final structure is binary with only two permittivity values \citep{Piggott2015, Osher1988, Sethian1999}. In the level set optimization, we first use the permittivity approach to update the permittivity continuously in each pixel until the figure of merit exceeds a threshold value. Then, we binarize the structure by using $(\epsilon + \epsilon_B)/2$ as the threshold.  The permittivity of every pixel is then changed to be either $\epsilon$ or $\epsilon_B$, depending on whether the existing permittivity is above or below the threshold. The binarized structure thus obtained is then further optimized using the level set technique. In general, the figure of merit achieved by the permittivity approach will be larger than that for level set methods, which is sensible given that optimizing the permittivity over a continuous range of values probes a much larger parameter space. 


\begin{figure}[H]
    \centering
    \includegraphics[width = 5.25 in]{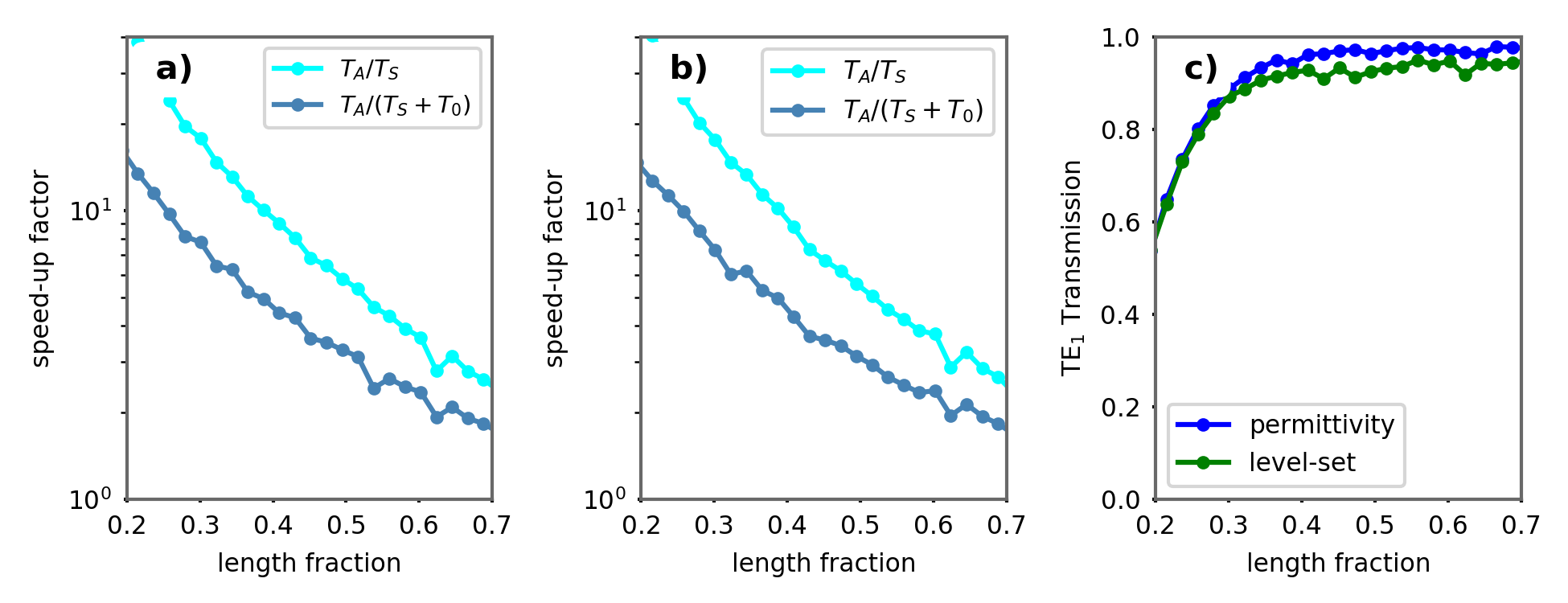}
    \caption{a)	The speed-up factors as a function of design region sizes for the permittivity optimization approach. The blue and cyan line are with or without the overhead time $T_0$, respectively. b) Same as a), except with the level set approach. c) The figure of merit for the optimized structure as a function of the design region size. The blue and green curves correspond to the permittivity and the level set approaches, respectively.}
    \label{fig:subdomain_scan}
\end{figure}

In Fig. \ref{fig:subdomain_scan}a) and b), we compare the speed-up factor, defined as $T_A/(T_0+T_S)$ for varying design region sizes for the two optimization approaches. The line segment for the figure-of-merit calculation is always outside the design region. The sizes are measured as the length fraction, which is the ratio of the design region side length to the side length of the total computational domain. A speed-up factor larger than one (including the overhead) indicates that solving the reduced system is more efficient than solving the original system. For both types of optimization approaches, solving the reduced system results in the substantial speed up. Even for relatively large design regions where the length fraction is almost 0.7 (in this example, this means the design region encompasses effectively the whole domain except the PML), which corresponds to  almost 50\% of the grid points in the domain being optimized, a speedup factor of two is still achievable. However, we observe that in this design example, a design region with a length fraction of about 0.4 is sufficient to achieve roughly the same performance in the optimized device.

In our plot, for each size of the design region we perform the gradient descent optimization only once. In practice, the number of optimization iterations may be a lot larger. For example, typically one will need to run the entire gradient descent optimization multiple times either for different initial structures, or to determine the optimal values for parameters such as the learning rate $\alpha$ in Eq. \eqref{eq:grad_update}. In the limit of a large number of optimization iterations, the overhead cost $T_0$ becomes negligible. It is instead more relevant to consider the speed-up factor per epoch, defined as $T_A/T_S$. In Fig. \ref{fig:subdomain_scan}a) and b) we also plot the speed-up factor per epoch as a function of design region sizes (in cyan), which gives a measure of the speed-up by our algorithm in the limit of large number of iterations.

In Fig. \ref{fig:subdomain_scan}a) and b), we see that the speed-up factor decreases as the design region increases. Certainly, the matrix $S$ for the reduced system should always have a dimension that is smaller than the matrix $A$ for the original system. The matrix $S$, however, is more dense. The speed of many linear system solvers are strongly influenced by both the dimension and the sparsity of the system matrix. As a simple estimate of the size of the design region beyond which point we can reduce our expectation of a significant speed up with our approach, we can calculate the size of the design region at which the number of non-zero elements of in the matrices $S$ and $A$ are the same. 

Given that our design region and our computational domain are square in geometry, let $N$ denote the number of grid points per side of the computational domain and $n$ the same for the design region. The matrix $A$ is a sparse matrix with $5N^2$ nonzero elements. The matrix $S$ contains two contributions to its sparsity pattern. One is from the $4n-2$ grid points on its boundary, which are densely coupled in $S$ and contribute $(4n-2)^2$ nonzero elements. The remaining $(n-2)^2$ grid points in the interior of the design region retain the same sparsity pattern as $A$ and contributes $5(n-2)^2$ nonzero elements to $S$, giving a total of $(4n-2)^2 +5(n-2)^2$ nonzero elements. Equating the number of nonzero elements for $S$ and $A$ gives a ratio $n/N \approx 0.5$ at which the nonzero in elements in $S$ and $A$ are approximately equal (though $S$ is still a factor of 4 smaller in dimension). In our simulation, at $n/N = 0.5$, the speed-up factor for both optimization approach is about $3$, which is indeed relatively low compared to smaller design regions.



In Fig. \ref{fig:subdomain_scan}c), for the optimal structure, we show the modal transmission coefficient from the TE$_0$ mode in the input waveguide to the TE$_1$ mode in the output waveguide as a function of design region size for both the permittivity approach (blue)  and the level-set approach (green). The optimal structures we obtained by solving either the original or the reduced system are the same,  which is expected since mathematically these two systems are equivalent to each other. For both optimization approaches, at small size of the design region (length fraction less than 0.4), the figure of merit of the optimal structures increases as the design region size increases. However, above this length fraction, the figure of merit saturates for the permittivity optimization, which suggests that the optimal design region for this problem lies in the regime where significant speed-ups can be achieved using our formalism. 



\begin{figure}[H]
    \centering
    \includegraphics[width = 5 in]{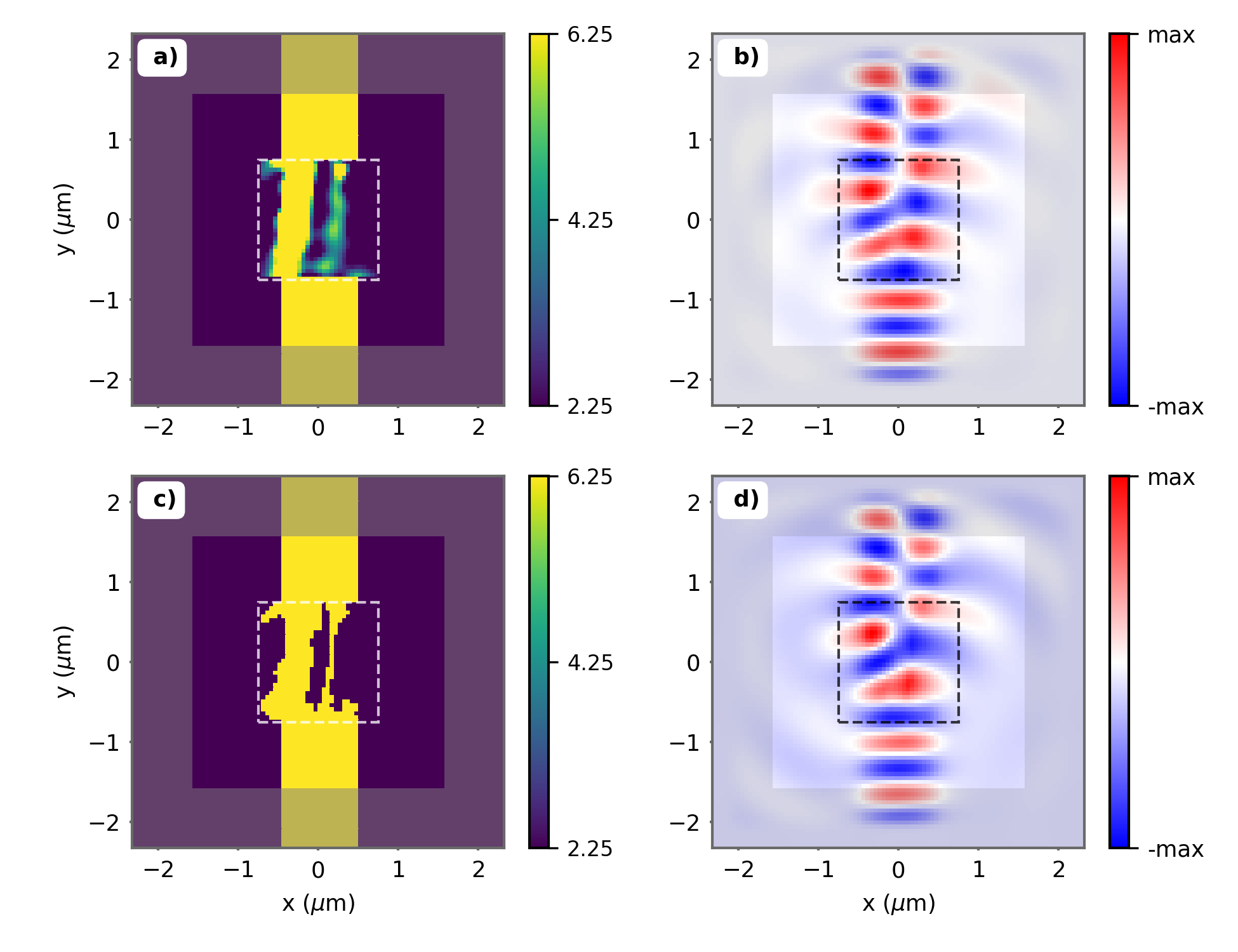}
    \caption{a)	Permittivity distribution, and b) electric field ($E_z$) distribution, for the optimized mode converter structure as obtained using the permittivity approach.  The design region, as denoted by the dashed white line, has a length fraction of $\approx 0.33$, correspond to a side length of approximaely 1.5 $\mu$ m. Regions of the PML are covered by a transparent gray layer. c) and d), same as a) and b), except with the use of the level set approach. }
    \label{fig:example_structures}
\end{figure}

We show, in Fig. \ref{fig:example_structures}a), the final optimized structure at the end of 450 iterations for a design region width that is 0.37 times the width of the total domain using adjoint optimization over a continuous range of permittivities and the same in Fig. \ref{fig:example_structures}c) with level set methods. As expected, the optimizing over a continuous range of permittivities results in a structure where the permittivity varies between 2.25 and 6.25 in the design region, whereas the level set method results in a binary structure. In Fig. \ref{fig:example_structures}b), we show the steady state Re$(E_z)$ field corresponding to the structure in Fig. \ref{fig:example_structures}a) and in Fig. \ref{fig:example_structures}d), we show the steady state Re$(E_z)$ field corresponding to the binary structure in Fig. \ref{fig:example_structures}c). In both cases, the optimal structure performs the expected functionality of mode conversion. 

\color{black}

\section{Conclusion}
In conclusion, we have developed a formalism based on the Schur complement to significantly reduce the computational degrees of freedom being simulated in each step of adjoint method based optimization of photonic devices. We have demonstrated the use of this algorithm by considering the optimization of a mode converter. The approach should be applicable in nanophotonic design in general. In addition, since the dimension of the reduced system can be a lot smaller than the original system, our approach may enable efficient local searches of the optimization landscape. 

This work was supported by the DOE ``Photonics at Thermodynamic Limits" Energy Frontier Research Center (Grant N\textsuperscript{\underline{o}} DE-SC0019140) and by an AFOSR MURI project (Grant N\textsuperscript{\underline{o}}. FA9550-17-1-0002).


\end{document}